\documentclass[10p,twocolumn]{article}

\usepackage{epsfig}
\usepackage{graphics}
\usepackage{graphicx}
\usepackage{dcolumn}
\usepackage{bm}

\begin{document}

\title{
Optical Nanoscopy of High-T$_c$ Cuprate Nano-Constriction Devices Patterned by Helium Ion Beams
}

\author{
A. Gozar$^{1,2*}$, N. E. Litombe$^{3,4}$, Jennifer E. Hoffman$^3$ and I. Bo\v{z}ovi\'{c}$^{1,2,4}$ \\ \\
$^1$Department of Applied Physics, Yale University, New Haven, Connecticut 06520, USA \\
$^2$Energy Sciences Institute, Yale University, West Haven, Connecticut 06516, USA \\
$^3$Department of Physics, Harvard University, Cambridge, Massachusetts 02138, USA \\
$^4$Brookhaven National Laboratory, Upton, NY 11973, USA
}

\maketitle

\textbf{Helium-ion beams (HIB) focused to sub-nanometer scales have emerged as powerful tools for high-resolution imaging as well as nano-scale lithography, ion milling or deposition. Quantifying irradiation effects is essential for reliable device fabrication but most of the depth profiling information is provided by computer simulations rather than experiment. Here, we use atomic force microscopy (AFM) combined with scanning near-field optical microscopy (SNOM) to provide three-dimensional (3D) dielectric characterization of high-temperature superconductor devices fabricated by HIB. By imaging the infrared dielectric response we find that amorphization caused by the nominally 0.5 nm HIB extends throughout the entire 26.5 nm thickness of the cuprate film and by $\sim$500 nm laterally. This unexpectedly widespread structural and electronic damage can be attributed to a Helium depth distribution substantially modified by internal device interfaces. Our study introduces AFM-SNOM as a quantitative nano-scale tomographic technique for non-invasive 3D characterization of irradiation damage in a wide variety of devices.}

In a pioneering recent work, superconductor-insulator-superconductor (SIS) Josephson Junctions (JJs) have been written by HIB in thin films of YBa$_2$Cu$_3$O$_7$ (YBCO) high-temperature superconducting (HTS) cuprates \cite{cybart15,cho15}. This is potentially quite exciting because a reproducible technology for fabricating high-quality and uniform arrays of in-plane SIS junctions remains the most desired goal in the field of HTS electronics. Such devices would allow manufacturing of sensitive magnetometers, voltage standards, and voltage-tunable THz radiation sources, and would provide a technology platform for high-speed computing \cite{fagaly06,clarke08,welp13,tafuri13}. SIS-JJ physics may also reveal the fingerprints of bosonic excitations involved in electron pairing, thus providing a clue about the as-yet unresolved mechanism of HTS in cuprates \cite{lee06,shim08}. However, HTS-based JJ devices typically exhibit properties of normal-metal weak links, while true SIS behavior is observed only very rarely. The challenge arises from the very short (1 - 2 nm) coherence length in HTS materials and the consequent sensitivity to point defects. Lithography by photons, electrons or heavier ions all exhibit excess damage or insufficient resolution for the required barrier sizes.

To verify the great potential of HIB for nano-fabrication \cite{ward06,hlawacek14,alkemade14}, one needs to study and evaluate irradiation-induced damage \cite{zeigler09,livengood09}. Compared to more commonly used heavy ions, Helium ions undergo lower energy loss per unit length, which leads to longer depth profiles and larger interaction volumes\cite{hlawacek14}. Since the defect density is largest near the stopping range of ions, HIB is expected to be less destructive for atomic layers closer to the surface \cite{hlawacek14,bell09}. However, the impact of the beam depends markedly on the softness of the materials, and back-scattering can occur from harder sub-surface layers \cite{bell09}. This is relevant for nanoscale devices based on multi-layer structures, which are typical of most practical electronic devices.

It is therefore essential to quantify the magnitude, lateral extent and depth profile of the irradiation damage in HIB-fabricated heterostructures. Existing knowledge about the interaction of ions with solids is obtained largely from computer trajectory simulations \cite{zeigler09}. Typical local tools like transmission electron microscopy or AFM may require extensive sample preparation or provide only limited information on sub-surface effects, chemical changes, or low-energy electronic and vibrational properties.

Here we introduce AFM-based SNOM as a powerful tool for non-invasive 3D nano-scale imaging of irradiated samples. The AFM tip provides topographical information, transfers sub-wavelength dielectric information into the far-field and increases optical sensitivity due to field enhancement \cite{keilman04,atkin12,govyadinov13,ni16,ni15}. The non-linear tip-sample interaction gives rise to higher harmonic content when the AFM tip operates in tapping mode. Crucially, different harmonics probe different sample depths so 3D dielectric information is intrinsically contained in every 2D AFM-SNOM scan \cite{govyadinov14,krutokhvostov12}. We use these properties to characterize HIB damage with a customized AFM-SNOM setup which was optimized for light access and tip/sample visualization (Fig.~1a).
\begin{figure}[t]
\centering
\includegraphics[scale=0.5]{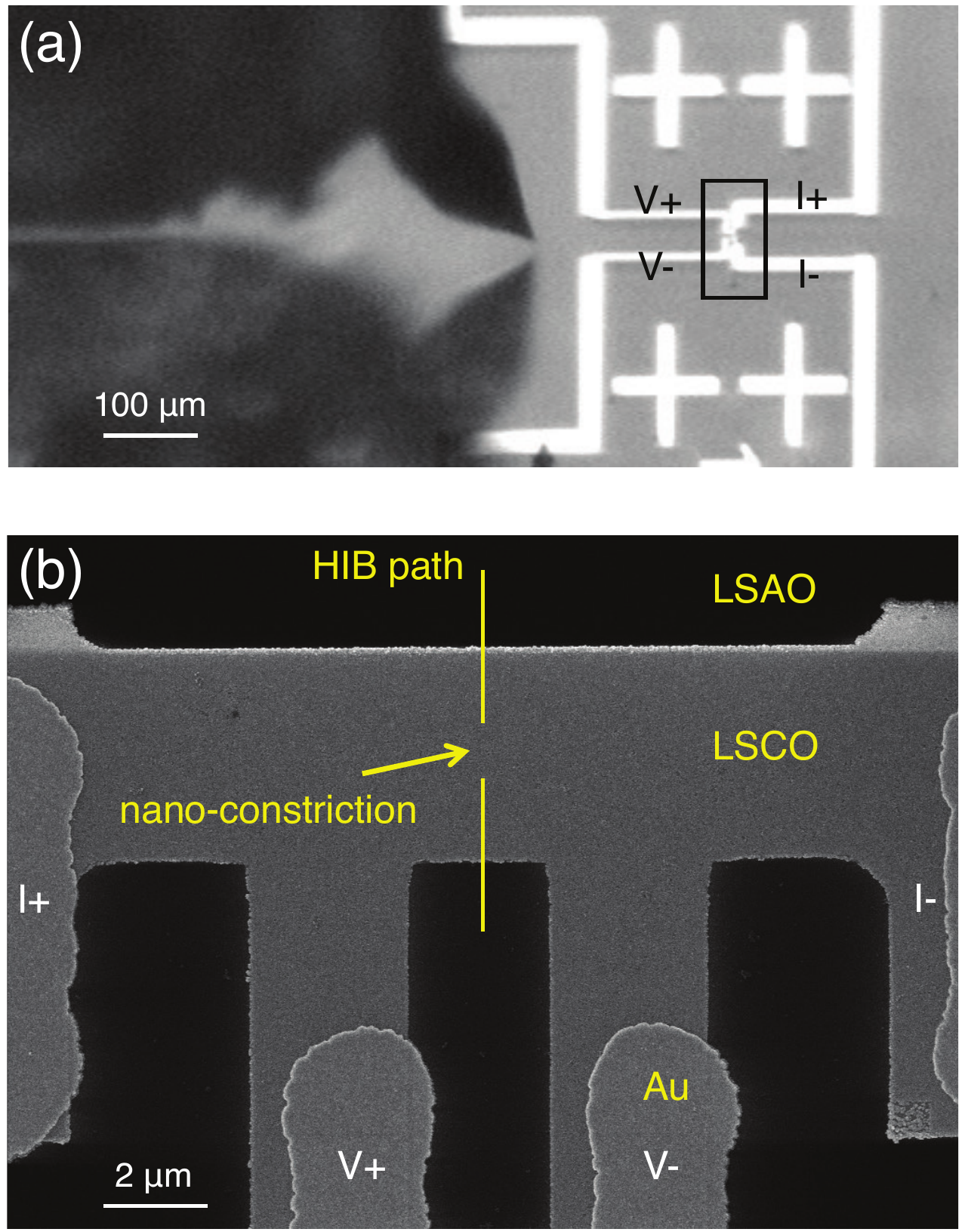}
\caption
{
Images of the AFM-SNOM setup and LSCO devices.  \textbf{a}. CCD image of the AFM tip acquired before approaching the device region, highlighted by the black rectangle. Visible are the 100 $\mu$m diameter etched W wire, its reflection in the LSAO substrate and the Au sputtered transport leads. \textbf{b}. He-ion microscope image of the device. Two vertical yellow lines show the nano-constriction patterned by the HIB. LSAO and LSCO stand for the substrate and film materials, respectively. V$+/-$ and I$+/-$ annotations in panel a and 90$^{\circ}$ rotated panel b correspond to the same Au leads.
}
\label{pdf1}
\end{figure}

Devices are fabricated from optimally doped La$_{1.84}$Sr$_{0.16}$CuO$_4$ (LSCO) superconducting films grown by molecular beam epitaxy (MBE) on LaSrAlO$_4$ (LSAO) substrates, see Methods. The films are 26.5 nm thick and atomically smooth with the critical temperature T$_c$ = 41 K. Each film was patterned by photolithography into 20 four-point-contact devices. Nano-constrictions were then written by HIB across the device bars \cite{litombe15}, see Fig.~1b. No changes in device resistance and T$_c$ were detected after photolithography or irradiation with doses  10$^{17}$ ions/cm$^2$. Although some narrow devices showed a reduced critical current, so far we were not able to observe true JJ behavior in magnetic fields or under microwave irradiation in our devices fabricated by HIB, irrespective of the irradiation dose or pattern geometry \cite{litombe15}. Our attention was focused on doses of 10$^{18}$ ions/cm$^2$ which ensure electrical isolation in our LSCO-based devices \cite{litombe15}, an order of magnitude higher than in YBCO \cite{cybart15,cho15}. This contrast may be related to the fact that superconductivity in YBCO is very sensitive to the structure of the 'soft' Cu-O chain layers, a peculiarity of the YBCO materials. 
\begin{figure}[t]
\centering
\includegraphics[scale=0.33]{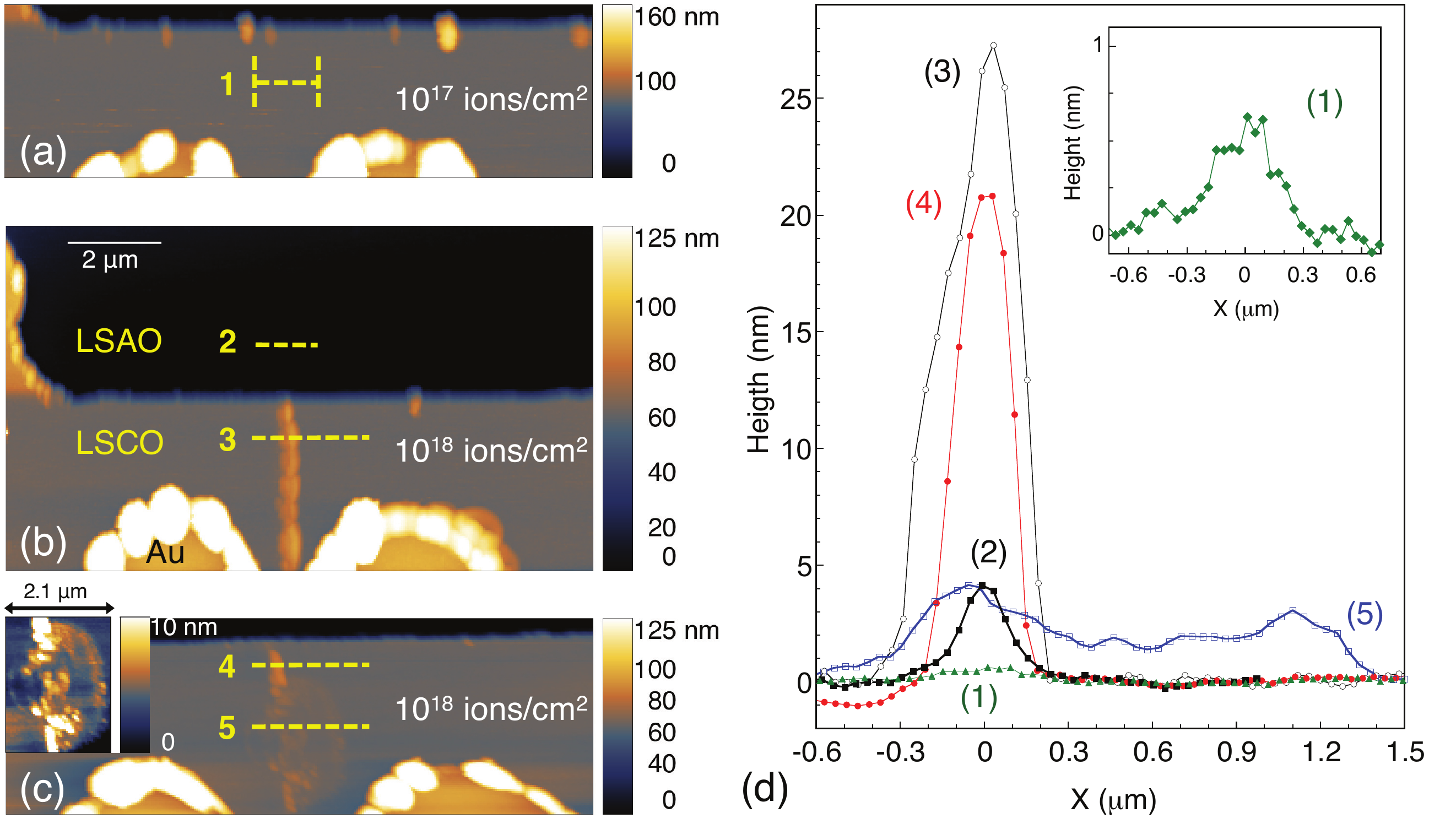}
\caption
{
AFM scans of the devices patterned by HIB.  \textbf{a}. A 12$\times$4.25 $\mu$m$^2$ scan of Device 1 (dose = $10^{17}$ ions/cm$^2$ and nominal nano-constriction gap = 200 nm).  \textbf{b}. A 12$\times$7 $\mu$m$^2$ scan of Device 2 (dose = $10^{18}$ ions/cm$^2$, nominal gap = 80 nm).  \textbf{c}. A 12$\times$3.5 $\mu$m$^2$ scan of Device 3 (dose = $10^{18}$ ions/cm$^2$, nominal gap = 800 nm). The inset of panel c shows a zoomed-in region of the constriction in Device 3. \textbf{d}. AFM line profiles across the HIB traces marked by yellow dashed lines in panels a-c with corresponding numerical labels. The inset magnifies trace (1) in panel d which corresponds to line 1 in panel a.
}
\label{pdf2}
\end{figure}

Figure~2 shows AFM data on the three representative devices studied in this work. The large area scans allow a direct comparison with the HIB microscope image shown in Fig.~1b. The apparent step height of 65 nm between the film and the substrate is larger than the film thickness (26.5 nm) because of deliberate over-etching of the film for a proper definition of the HTS bridge structure. Representative traces on the samples are marked by dotted lines and topographic profiles are shown in Fig.~2d. For the lower dose of 10$^{17}$ ions/cm$^2$, Fig.~2a, we averaged the height profile over multiple consecutive scan lines to search for subtle effects of irradiation on LSCO topography, as shown by the larger ticks of line $\#$1 in Fig.~2a. The corresponding data, green line in the inset of Fig.~2d, show swelling at the surface of about 5 \AA. We detected no HIB-induced damage on the LSAO substrate at 10$^{17}$ ions/cm$^2$.

\begin{figure}[t]
\centering
\includegraphics[scale=0.42]{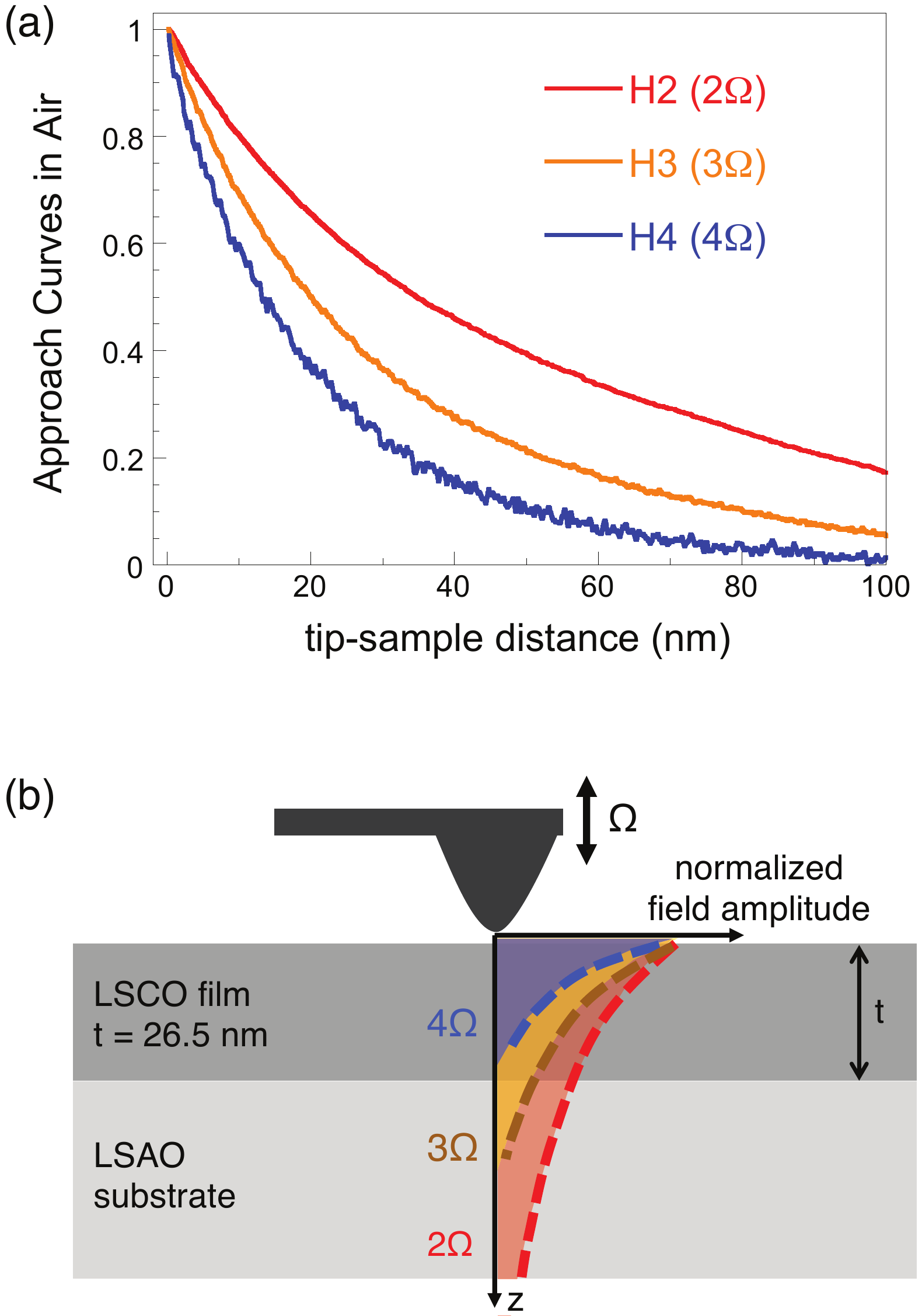}
\caption
{
Principles of nano-optical tomography.  a. Normalized SNOM signal as a function of tip-sample distance while approaching in air a thick LSCO sample for the $2^{nd}$ (H2), $3^{rd}$ (H3) and $4^{th}$ (H4) harmonics of the AFM tapping frequency $\Omega$ = 28.1 kHz. The range of interactions estimated from the height at which the signal decreases below 10\% of the maximum value are 120 nm, 80 nm and 50 nm for H2, H3 and H4 respectively. The near-field probing range inside our devices is reduced by more than a factor of two, see Methods. b. Rendition of how the dielectric depth information is encoded in the mid-infrared optical signal demodulated at the $2^{nd}$, $3^{rd}$ and $4^{th}$ harmonics. The estimated depth profile for each harmonic is indicated in relation to the actual LSCO film thickness of 26.5 nm.
}
\label{pdf3}
\end{figure}
At a larger dose of 10$^{18}$ ions/cm$^2$, the HIB beam traces become clearly visible in Device 2, Fig. 2b. The LSCO film swells by $\sim$25 nm, comparable to the film thickness (profiles $\#$3 and $\#$4). In the LSAO substrate, swelling is a factor of five less than in the cuprate. The beam damage footprint as inferred from the swelling in the AFM topography is about 500 nm. Given that the He ion beam, although focused to sub-nanometer size, has an effect spreading over such a large area we do not expect to observe lithographically defined features with a size smaller than a few hundreds of nanometers. Indeed, the nominally 80 nm large nano-constriction gap remains invisible in Fig. 2b, although 80 nm is larger than the AFM topography resolution, which is given by the tip apex radius of about 10-20 nm. Even in Device 3, the damage is observed to extend more throughout and perpendicular to the nominally 800 nm gap, see Fig. 2c.

Our data demonstrate substantial damage at the threshold dose of 10$^{18}$ ions/cm$2$. While the lateral straggle of $\sim$500 nm due to the HIB appears to be independent of material, the observed height difference between the irradiated substrate (line $\#$2) and film (lines $\#$3 and $\#$4) is dramatic. The difference cannot be attributed to the beam geometry because the depth of focus extends well beyond the film thickness (26.5 nm). On the other hand, for a homogeneous material we expect top surface layers to be less affected than the 'bulk' due to the large depth profile of He beams. The data thus indicate that the dramatic topographic contrast is due to enhanced backscattering in the LSCO film from the irradiation-harder LSAO substrate. Sample swelling is then caused by the formation and accumulation of He nano-bubbles at the film-substrate interface.
\begin{figure}[t]
\centering
\includegraphics[scale=0.4]{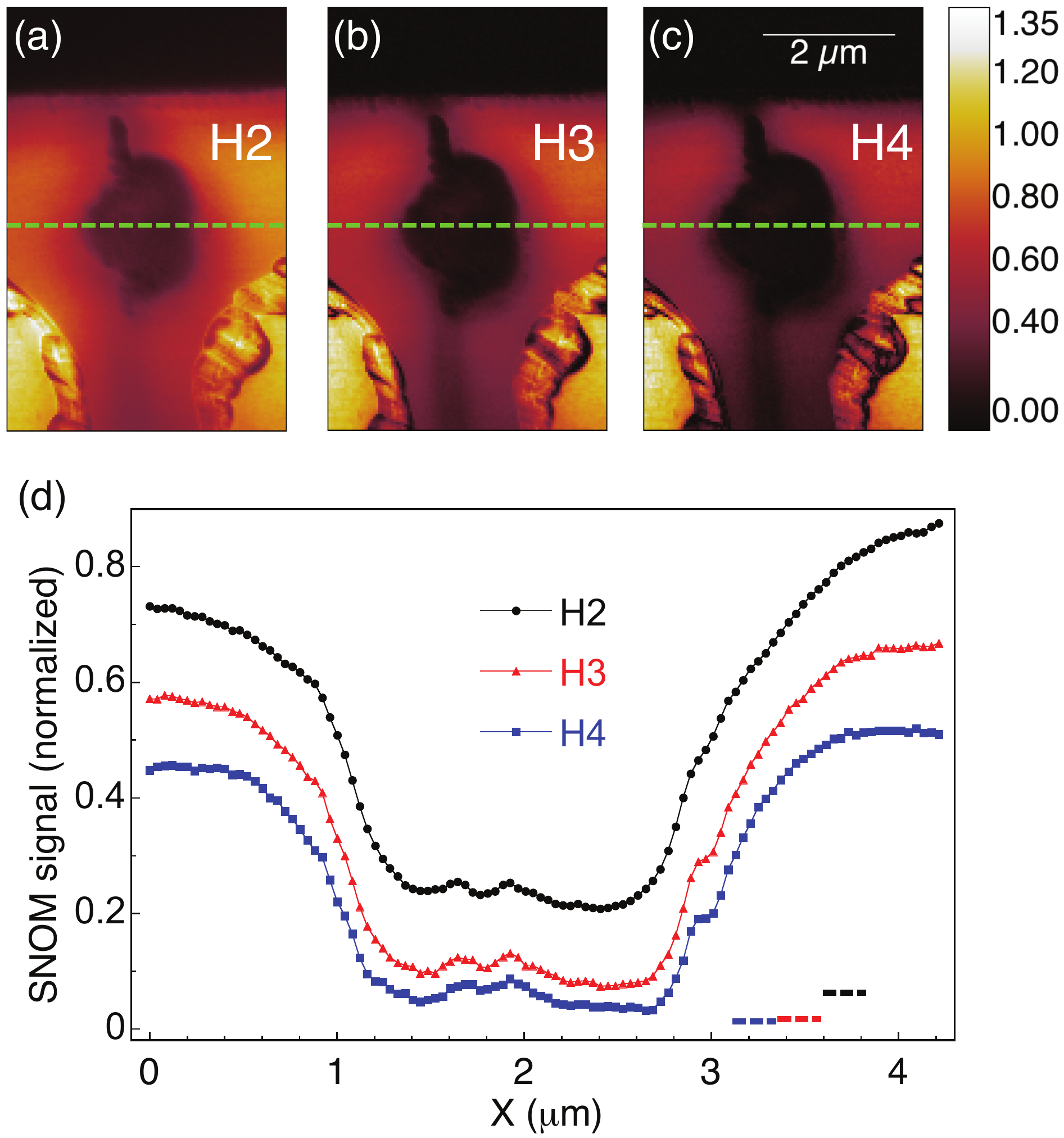}
\caption
{
Dielectric tomography of irradiated sample: data taken from Device 3. SNOM images show the depth dependence of the damage in the nano-constriction area that was shown topographically in Fig.~2c. The SNOM signal taken with  = 10.6 m is demodulated at the $2^{nd}$ (H2), $3^{rd}$ (H3) and $4^{th}$ (H4) harmonic of the tapping frequency in panels \textbf{a}, \textbf{b} and \textbf{c} respectively. The values are normalized to the near-field signal on the Au pads. \textbf{d}. Line profiles corresponding to the traces shown in panels a-c. The black, red and blue lines correspond to H2, H3 and H4, respectively. The near-field signal on the LSAO substrate is shown for reference on the bottom right, demonstrating that the HIB damage renders LSCO strongly insulating.
}
\label{pdf4}
\end{figure}

Figures~3 and 4 demonstrate the complementary use of AFM-SNOM to provide 3D dielectric characterization. In Fig.~3 we illustrate the principle of nano-optical tomography. With the help of approach curves in air (Fig.~3a) and model calculations (see Methods) we are able to estimate the actual probing depth of the different harmonic components of the SNOM signal (denoted by H2, H3 and H4 in Fig.~3b) into the LSCO sample. We notice that while H3 and H4 signals originate almost entirely from the top layer, H2 appears to have a substantial contribution from the substrate.

Fig.~4a-c shows AFM-SNOM data at  = 10.6 $\mu$m in Device 3 for several harmonics of the tapping frequency. The increased surface sensitivity at higher harmonics can be observed in the data: the dielectric contrast between pristine LSCO and in the irradiated areas becomes less visible with decreasing harmonic order. The HIB dielectric footprint as revealed by SNOM scales well with the morphological one inferred from the AFM data. In the constriction area, H2 decreases by about 70\% while H3 and H4 decrease by about 85\%, compared to the intensity from the non-irradiated film. In spite of the different topography between the written line and constriction area (lines $\#$4 and $\#$5, respectively in Fig.~2c) the infrared response in both of these regions is very similar to each other and also to the signal collected from the highly insulating LSAO substrate. The near-field data clearly show that irradiation effects at doses of 10$^{18}$ ions/cm$^2$ have a large impact on the electronic properties of the HTS films, optically turning it into an insulator over an area exceeding the HIB spot size by several orders of magnitude.

We performed a quantitative analysis of the AFM-SNOM data by analyzing the tip-sample interaction within the dipole model \cite{mcleod14,jones10,raschke03,cvitkovic07,ocelic06}. We verified that the model is applicable in this case by several self-consistency checks, see Methods. The numerical results fully support the idea of complete film amorphization in the irradiated and constriction areas. Using LSCO and Au, respectively, as reference materials \cite{uchida91,olmon12,chen04}, the $3^{rd}$ and $4^{th}$ harmonic data from the damaged regions yield the permittivity $\varepsilon_i \approx$ 1.3-2.5 at  = 10.6 $\mu$m.  The actual value could be 2-3 times larger due to the formation of He bubbles and the associated inhomogeneous structure. The inferred value of i, very close that in LSAO substrate at the same wavelength (LSAO ≈ 4.1 at  = 10.6 $\mu$m) , validates our conjecture about a strongly insulating behavior. Data at different harmonics also confirm that the depth of the damage spans the entire cuprate film thickness, see Methods.

Our findings should be considered when applying HIB to nano-scale lithography, and explored systematically for a given material-substrate system. By providing tomographic optical information with nm-scale resolution in a non-destructive way, we introduce AFM-SNOM as an excellent new tool for evaluation of lithography in complex hetero-epitaxial structures. We show that the promising technique of HIB lithography must take into account the complexities of backscattering in layered systems. By recording a wider range of fundamental frequencies this technique could be further refined to provide higher resolution of depth profiling for thinner heterostructures.

\paragraph{Acknowledgements}

\noindent

This research was supported by the U.S. Department of Energy, Basic Energy Sciences, Materials Sciences and Engineering Division. X.H. and A.G. are supported by the Gordon and Betty Moore Foundation’s EPiQS Initiative through Grant GBMF4410. A.G. also acknowledges support from the DOE Early Career Research Program, Grant No. 2005410. We would like to thank C. Huyn from Carl Zeiss for help in HIB device patterning.

\paragraph{Author contributions}

\noindent

I.B. and A.G. designed the experiment. N.E.L. fabricated the samples. A.G. built the near-field setup, carried out the measurements and analyzed the data. All authors discussed the results and commented on the manuscript.

\paragraph{Competing financial interests}

\noindent

The authors declare no competing financial interests.

\paragraph{Correspondence}

\noindent

$^*$ adrian.gozar$@$yale.edu

\paragraph{Supporting Information}

\noindent

\smallskip

\textbf{A. Synthesis}

Optimally-doped, 26.5 nm thick LSCO films were grown on insulating LSAO substrates polished perpendicular to the crystallographic [001] direction. For synthesis, we used atomic-layer-by-layer molecular beam epitaxy (ALL-MBE) technique \cite{bozovic01,gozar08,logvenov09,sochnikov10,gozar16}. The stoichiometry was controlled before and during deposition by a scanning quartz crystal monitor and by a calibrated custom-built 16-channel atomic absorption spectroscopy system, respectively. The film structure was characterized in real time by reflection high-energy electron diffraction (RHEED) and ex-situ by X-ray diffraction and AFM. The films were found to be atomically smooth and with no secondary phase precipitates. Superconducting properties and a critical temperature T$_c$ = 41~K were measured on the entire $10 \times 10$ mm$^2$ film by contactless mutual inductance technique in transmission geometry.

\textbf{B. Lithography}

Each sample was patterned by photolithography into 20 four-point-contact devices. The individual devices (Fig.~1) are 4~$\mu$m wide; the current leads are separated by 15~$\mu$m and the voltage leads by 2.5~$\mu$m. The leads were made out of the same LSCO material as the bridge, with Au contacts sputtered atop. The superconducting properties were unchanged after these patterning steps.

\textbf{C. HIB writing}

A Carl Zeiss Orion Helium Ion Microscope equipped with a nano-patterning engine was used for writing and imaging the devices \cite{litombe15,scipioni08}. In order to minimize heating and charging effects, we used a 30 kV accelerating voltage and low beam currents (1 pA). Drift and charging were minimized by sample grounding and by using an electron flood gun operated at 500 V with the current of 2-3~$\mu$A. The chamber pressure was maintained at p = $5 \times 10^{-7}$ Torr. The HIB was focused to 0.5 nm spot size. The lines were written as a predefined linear array of spots 0.25 nm apart. We measured three nano-constriction devices with the doses and nominal gaps as follows: 10$^{17}$ ions/cm$^2$ and 200 nm gap for Device 1, 10$^{18}$ ions/cm$^2$ and 80 nm gap for Device 2, 10$^{18}$ ions/cm$^2$ and 800 nm gap for Device 3.

\textbf{D. AFM-SNOM measurements}

Near-field and AFM data were acquired simultaneously in a custom-built AFM-SNOM setup. The sample chamber was optimized for laser light access to the AFM tip and for sample visualization (Fig.~1a). AFM tips are reproducibly obtained by etching 100~$\mu$m diameter W wires in a 2M-NaOH solution \cite{klein97}. The typical tip radii are in the 20 – 25 nm range, values which also define the lateral resolution of our AFM. The tips are glued to a piezo-actuated quartz tuning fork (TF) resonator operated at its resonance frequency $\Omega$ = 28.1 kHz. AFM data were taken in amplitude-modulation mode with the feedback based on the TF piezocurrent. The AFM oscillation amplitude was A = 80 nm. Light from a grating tunable CO$_2$ laser with a wavelength  = 10.6~$\mu$m was focused on the tip with a the help of an off-axis Au-coated copper paraboloid. We used a backscattering geometry and a homodyne setup to amplify the near-field signal. The optical signal was focused on a LN$_2$-cooled Mercury Cadmium Telluride detector and demodulated in real-time up to the $4^{th}$ harmonic of the tapping frequency . Background scattering contribution at the $2^{nd}$ harmonic (H2) did not exceed 10\% of the signal and for the $3^{rd}$ and $4^{th}$ harmonics (H3, H4) it was completely suppressed.

\textbf{E. Data analysis}

In the SNOM images of Fig.~3 we analyze the tip-sample interaction within the dipole model \cite{mcleod14,jones10,raschke03,cvitkovic07,ocelic06,ocelic07}, approximating the metallic AFM tip with a sphere (of radius R) whose polarizability is modulated by the electrostatic interaction with a dielectric medium, placed at a distance $z(t)$ along the $z$-axis. 

The scattered field is dependent on the effective tip polarizability given by:
\begin{equation}
\alpha_{eff} = \frac{ (1 + r_p)^2 \alpha_t}{1 - \beta(\varepsilon) \alpha_t / [16 \pi \varepsilon_0 (R + z_0 + z(t))^3]}
\label{e1}
\end{equation}
where $\varepsilon_t = 4 \pi \varepsilon_0 R^3 (\varepsilon_t - 1)/(\varepsilon_t + 2)$ is the ‘bare’ sphere polarizability, $\varepsilon_t$ is the complex dielectric constant of the AFM tip material $\beta (\varepsilon) = (\varepsilon - 1) / (\varepsilon + 1)$, $\varepsilon$ is the spatially varying dielectric constant of the sample, $z_0$ is the tip-height offset, and $z(t) = A[1 + \cos(t)]$ is the harmonic time-dependent distance between the tip and the sample in the tapping mode. With $r_p$ the far-field Fresnel coefficient for p-polarized light, the factor $(1 + r_p)^2$ takes into account the contribution of light reflected from the sample to both the tip polarizability and the scattered radiation \cite{cvitkovic07}. The dependence of $\alpha_{eff} = \alpha_{eff} (\omega)$ on the incident light frequency is implicit through $\varepsilon, \varepsilon_ t, r_p$ and $\beta$. The near-field intensity demodulated at $n \Omega$ from material '\emph{(x)}' is given in our configuration by:
\begin{equation}
I_n^{(x)} = C_1 \cdot \sigma_n^{(x)} \cdot \vert \cos ( \varphi_n^{(x)} - \psi ) \vert
\label{e2}
\end{equation}
where $C_1$ is a proportionality constant taking into account the overall system response, $\sigma_n e^{i \varphi_n}$ is the ‘n-th’ Fourier amplitude of the effective polarizability $\alpha_{eff}$ from Eq.~\ref{e1}, and $\Psi$ is an arbitrary and constant phase \cite{ocelic06}.

Our experimental parameters allow us to bring Eq. (1) to a simplified analytical form. The denominator of this equation can be written as $1 - f(\varepsilon,\varepsilon_t) \xi^3 / [b + \cos(\Omega t)]^3$ where by definition $f(\varepsilon, \varepsilon_t) \equiv 0.25 (\varepsilon - 1) (\varepsilon_t - 1) / [(\varepsilon + 1) (\varepsilon_t + 2)]$, $\xi = R / A$ and $b = 1 + (R + z_0) / A > 1$. At the wavelength  $\lambda = 10.6 \mu$m ($\hbar \omega \simeq 125$~meV), the dielectric functions are $\varepsilon_t \simeq -2144 + i1015$ for the tungsten tip \cite{palik98}, $\varepsilon_{Au} \simeq -4679 + i1674$ for gold \cite{olmon12}, $\varepsilon_{LSCO} \simeq -16 + i46$ for the cuprate film \cite{uchida91} and $\varepsilon_s \simeq 4.1$ for the LSAO substrate \cite{chen04}. These numbers imply an absolute value of $f(\varepsilon, \varepsilon_t)$ close to unity. At the same time, given that our tapping amplitude is $A \simeq 80$~nm, the effective tip radius is $R \simeq 25$~nm, and the height offset $z_0$ is close to zero if the AFM tip is touching the sample at the point of closest approach, we have $\xi^3 / [b + cos(\Omega t)]^3 < \xi^3 / b^3 \approx 0.008 \ll 1$. As a result, retaining only the 1$^{st}$ term in the Taylor expansion of the denominator is a very good approximation. From Eq.~\ref{e1}, the end result for the complex near-field amplitude at the frequency $n \Omega$ for the material '\emph{(x)}' is given by:
\begin{equation}
\sigma_n e^{i \varphi_n} = (1 + r_p)^2 \left( \frac{\varepsilon_t - 1}{\varepsilon_t + 2} \right)^2 \left( \frac{\varepsilon - 1}{\varepsilon + 1} \right) \int_0^{2 \pi} dx  \frac{\cos nx}{(b + \cos x)^3}
\label{e3}
\end{equation}
For simplicity, we denote the value of the last integral in the equation above by $g(n,b)$. We can see that in this approximation the intensity $I_n^{(x)}$ in Eq. (2) depends on the order of the harmonic only through $g(n,b)$. The dependence on the tip-sample distance is implicit through the dependence on the parameter $b$. One can easily check by plotting $g(n,b)$ that this approximation already captures qualitatively the idea behind the 3D optical nano-tomography: the higher the harmonic, the smaller the near-field interaction range, and accordingly the volume of the sample probed in SNOM experiments. 

Several consistency checks prove the validity of the dipole model used here. There are two predictions that follow immediately from the above relations. The first is that $I_n^{(x)} / I_p^{(x)} = g(n,b) / g(p,b)$. In other words, the ratio of the near-field intensities at different harmonics should not depend on '\emph{(x)}', i.e. on the material under study. Experimentally, we find that $I_3^{(x)} / I_4^{(x)}$ is equal to 1.75, 2.07 and 2.1 for $x =$~Au, LSCO and LSAO, respectively, which is almost independent on $x$. Moreover, these values are reasonably close to the theoretical value $g(3,b)/g(4,b) = 1.54$ where $b = 1 + R/A \simeq 1.3$ for the tapping amplitude $A \simeq 80$~nm and the tip radius $R \simeq 25$~nm (as seen from high-resolution SEM images). Note that the agreement between theory and experiment can be made arbitrarily good if we use the tip radius as an effective fitting parameter (in fact, this ‘inverse’ procedure renders $b \simeq 1.75$ and correspondingly $R_{eff} \approx 60$~nm, not far from the real value). This comparison is expected to become less good at lower harmonics, i.e. H2, as they can pick some contribution from the background scattering as well as from the substrate, see Fig.~3. For the 2$^{nd}$ and 3$^{rd}$ harmonics, $I_2^{(x)} / I_3^{(x)} = 2.01$ and 2.47 for $x =$ Au and LSCO and, respectively, which is not too far from the theoretical value $g(2,b) / g(3,b) = 1.44$. This agreement independently attests that the background contribution to H2 is relatively small. For $x = $ LSAO, we get $I_2^{(x)} / I_3^{(x)} = 7.45$; this discrepancy is likely due to the smaller signal obtained from the highly insulating LSAO.

The second consistency check refers to the prediction of Eqs.~(\ref{e2}) and (\ref{e3}) is that $I_n^{(x1)} / I_n^{(x2)}$ should not depend on the harmonic order '\emph{n}'. This is because the contribution from $g(n,b)$ cancels out in the ratio, and also because for the above values of $(x)$ for $x =$~Au, LSCO and LSAO the value of the phase $\varphi_n$ in Eq.~(\ref{e3}) is found to be almost vanishingly small, rendering $I_n^{(x)} \approx \sigma_n^{(x)} \cos(\Psi)$ in Eq. (\ref{e2}). This is again verified by the experiment for the Au/LSCO pair as we obtain $I_n^{(Au)} / I_n^{(LSCO)} =$~1.31, 1.61 and 1.92 for $n =$~2, 3 and 4, respectively. These numbers again compare reasonably well with the theoretical value of 2.24 obtained from the above values for the dielectric constants. As expected, due to the smaller signal we observe some discrepancies for the LSAO sample. Overall, the experimental validation of the two independent consistency checks provided by Eqs.~(\ref{e2}) and (\ref{e3}) indicates that they are reasonable starting points for our analysis.
\begin{figure}[t]
\centering
\includegraphics[scale=0.43]{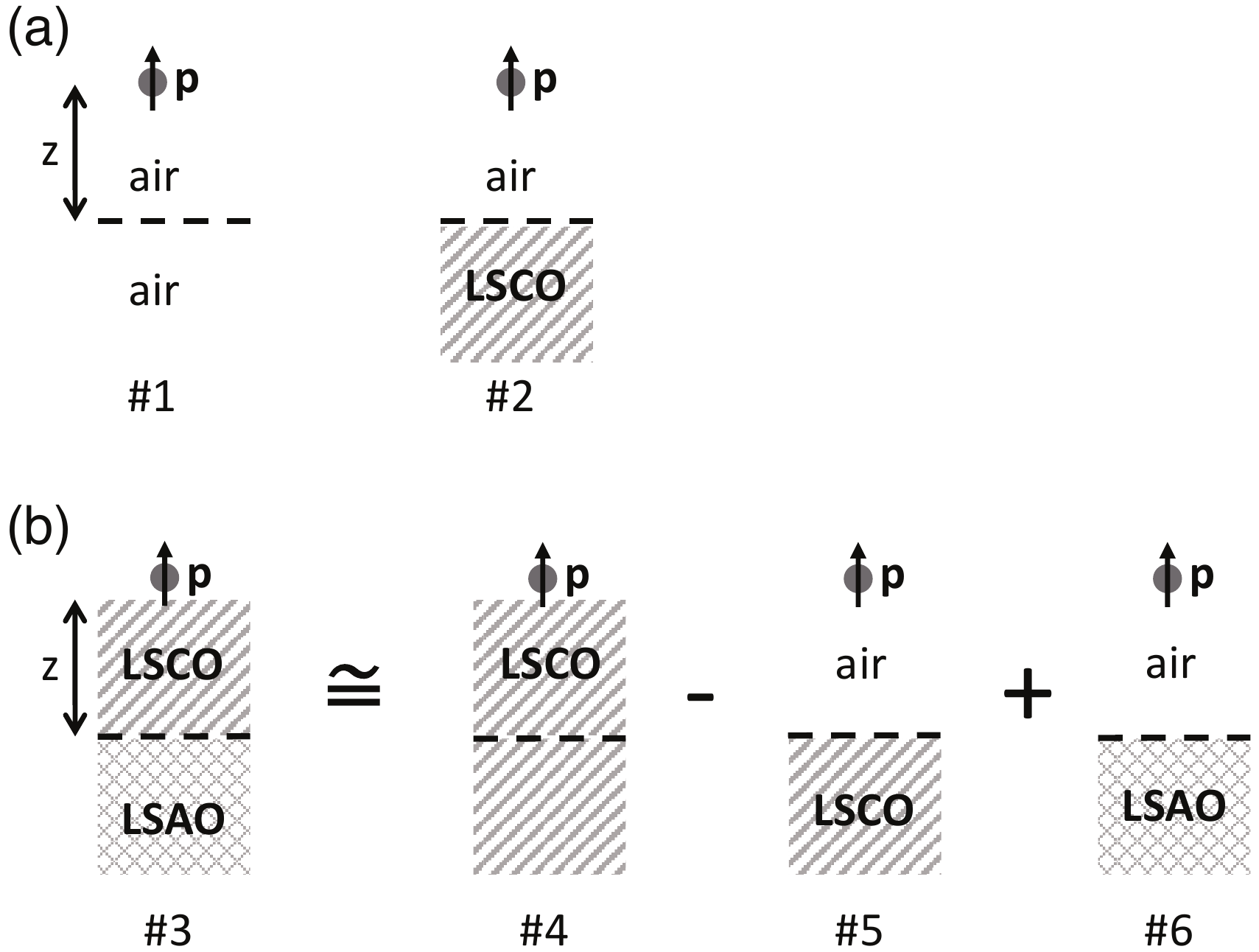}
\caption
{
\textbf{Estimation of the probing range of near-fields in the sample.}  Approach curves in air as the ones shown in Fig.~3 measure the difference in the detector signal between configurations like \#1 and \#2 as a function of tip-sample separation, see panel \textbf{a}. The AFM tip is depicted schematically by a sphere with a dipole moment \textbf{p}. The probing range inside the sample can be inferred from estimating how far below the LSCO sample should a putative LSAO substrate be located so that the near fields can still detect its presence. Equivalently, we want to determine what is the maximum LSCO film thickness so that the difference between configurations \#3 and \#4 is measurable, see panel \textbf{b}. In the 1$^{st}$ order, this difference is equal to the difference between configurations \#5 and \#6 which can be evaluated using Eq.~\ref{e3}, see text in the Methods section.
}
\label{pdf5}
\end{figure}

With the help of Eq.~(\ref{e3}), we can estimate the effective dielectric constant of the irradiated area $\varepsilon_i$ by normalization to the corresponding signal from ‘reference’ regions like gold or the cuprate film. The data from the 3$^{rd}$ and 4$^{th}$ harmonics render at $\lambda = 10 \mu$m ($\hbar \omega \simeq 125$~meV) $\varepsilon_i \approx 1.32$ if we use LSCO and $\varepsilon_i \approx 2.5$ using Au, respectively, as reference materials. Given the fact that LSCO and Au are quite different in nature, and taking into account the error bars in the estimates of their dielectric constants, the fact that we get similar results for the permittivity of the damaged area attests again that our approximations are reasonable. The small value of $\varepsilon_i$ that we obtain fully confirms our initial speculation that irradiation leads to a complete film amorphization. Data at multiple harmonics should also allow for the determination of the thickness of the irradiated layer. In the first order, we find no solution for an assumed damaged layer in the interval from zero to 26.5 nm, the LSCO film thickness. This physically means that the irradiation damage occurs throughout the entire cuprate film. We point out that the obtained values for $\varepsilon_i$ are lower bounds. Topography data show that the surface of the film swells sizably in the patterned regions, presumably due to the formation of trapped bubbles \cite{livengood09}. As a result, we are dealing with an inhomogeneous sample and an effective medium approach \cite{bruggeman35} should be more accurate. In principle this could increase by 2 - 3 times the true value of $\varepsilon_i$. A systematic study of samples with various thicknesses and irradiation levels where such subtle effects are included is the scope of future work.

\textbf{F. Probing depth of SNOM signal demodulated at different harmonics of the tapping frequency}
 
We can introduce an operational definition of the range probed by each harmonic for a given pair of materials, one acting as the sample and the other as the medium in which the sample is located. We define this range as the maximum distance at which the presence of a sample placed in a medium in the near-field of the AFM tip can still be detected. For example, the approach curves in Fig. 3a tell us the ranges at which a thick LSCO material is detected by each harmonic when we approach it in air. Choosing an operational cut-off criterion as a drop below 10\% of the maximum signal (obtained when the tip is touching down on the sample) the ranges of H2, H3 and H4 are about 120, 80 and 50 nm respectively.

A pictorial definition equivalent to the one above involves the determination of the maximum distance at which a difference between configurations like \#1 and \#2 in Fig.~5a can still be measured, i.e. being able to distinguish between the case when the sample is in the proximity or very far away from the AFM tip. This probing range is expected to be material dependent because of screening. In the particular case of a thin LSCO film on a very thick LSAO substrate, we ask how close to the cuprate surface does the LSAO have to be so that its presence is detectable. This means a measurable difference between configurations like \#3 and \#4 in Fig.~1b. We assume the tip to be as close as possible to the LSCO surface for achieving the maximum sub-surface sensitivity. In 1$^{st}$ order, i.e. neglecting multiple scattering, this difference is the same as that between configurations \#5 and \#6. This identification is helpful because we recognize that at this level of approximation the difference between \#5 and \#6 can be evaluated in a straightforward manner using the relations already derived in the previous data analysis section.

The tip-sample distance appears thorough the parameter $b = 1 + (R + z_0) / A$ in the function $g(n,b)$ of Eq.~\ref{e3}. We denote by $b_{max} = 1 + (R + z_{0,max}) / A$ the interaction range of a given harmonic as defined above. Knowing $b_{max}$ for the air - LSCO pair from approach curves like the ones shown in Fig.~3 and using Eq.~\ref{e3} we can obtain $b_{max}$ for (hetero-)structures involving other configurations. For LSCO grown on a LSAO, our case, we have:

\begin{eqnarray}
 \left \vert \frac{\varepsilon^{LSCO} - 1}{\varepsilon^{LSCO} + 1} \right \vert \cdot g(n, b_{max}^{air}) 
\simeq
\nonumber
\\
\left ( \left \vert \frac{\varepsilon^{LSCO} - 1}{\varepsilon^{LSCO} + 1} - \right \vert -  \left \vert \frac{\varepsilon^{LSAO} - 1}{\varepsilon^{LSAO} + 1} \right \vert \right) \cdot g(n, b_{max}^{LSCO})
\label{e4}
\end{eqnarray}
We know $b_{max}^{air}$ from Fig.~3a so from Eq.~\ref{e4} we can determine $b_{max}^{LSCO}$ which, in 1$^{st}$ order, represents the maximum thickness of LSCO which still allows probing of the LSAO substrate. Equivalently, this is the range of near-field interactions for the SNOM signal demodulated at $n \Omega$ in our structure. Using the values of the dielectric functions for LSCO and LSAO substrate quoted in the previous section ($\varepsilon_{LSCO} \simeq -16 + i 46$ and $\varepsilon_s \simeq 4.1$) and the values of $b_{max}^{air}$ inferred from Fig.~3a ($b_{max}^{air} \approx $ 120 nm, 80 nm and 50 nm for H2, H3 and H4 respectively) we obtain $b_{max}^{LSCO} \approx$  60 nm, 35 nm and 16 nm for H2, H3 and H4 respectively. The values are about half of $b_{max}^{air}$ and form the basis of Fig.~3b where we depict the probing depth of each harmonic in relation to the thickness of the cuprate film in our devices.

\end{document}